\documentclass[prb,twocolumn,showpacs,preprintnumbers,amsmath,amssymb]{revtex4}
\usepackage{graphicx}
\usepackage{dcolumn}
\usepackage{bm}

\preprint{submitted to Applied Physics Letters}

\begin{document}

\title{3\textit{d}-electron induced magnetic phase transition in
half-metallic semi-Heusler alloys}

\author{I. Galanakis}\email{galanakis@upatras.gr}
\affiliation{Department of Materials Science, School of Natural
Sciences, University of Patras,  GR-26504 Patra, Greece}

\author{E. \c{S}a\c{s}{\i}o\u{g}lu}\email{e.sasioglu@fz-juelich.de}
\affiliation{Institut f\"{u}r Festk\"{o}rperforschung,
Forschungszentrum J\"{u}lich, D-52425 J\"{u}lich, Germany  \\ and
Fatih University,  Physics Department, TR-34500, B\"{u}y\"{u}k\c
cekmece, \.{I}stanbul, Turkey}

\author{K. \"{O}zdo\u{g}an}\email{kozdogan@gyte.edu.tr}
\affiliation{Department of Physics, Gebze Institute of Technology,
Gebze, TR-41400, Kocaeli, Turkey}

\date{\today}

\begin{abstract}
We study the effect of the non-magnetic 3\textit{d} atoms on the
magnetic properties of the half-metallic (HM) semi-Heusler alloys
Co$_{1-x}$Cu$_{x}$MnSb and Ni$_{1-x}$Cu$_{x}$MnSb ($0 \leq x \leq
1$) using first-principles calculations. We determine the magnetic
phase diagram of both systems at zero temperature and obtain a
phase transition from a ferromagnetic to an antiferromagnetic
state. For low Cu concentrations the ferromagnetic RKKY-like
exchange mechanism is dominating, while the antiferromagnetic
superexchange coupling becomes important for larger Cu content
leading to the observed magnetic phase transition. A strong
dependence of the magnetism in both systems on the position of the
Fermi level within the  HM gap is obtained. Obtained results are
in good agreement with the available experimental data.
\end{abstract}


\maketitle

In half-metallic semi-Heusler alloys with the chemical formula
XMnZ, where X is a high-valent transition metal atom and Z a
\textit{sp}-element, the magnetization is usually confined to the
Mn sublattice and the total magnetic moment assumes integer values
given by the Slater-Pauling rule.\cite{GalanakisFull}
Additionally, the Mn-Mn distance is rather large and thus the
3\textit{d} states belonging to different Mn atoms do not overlap
considerably. The ferromagnetism of the Mn moments stems from an
indirect exchange interaction mediated by the conduction
electrons. Therefore, the magnetic properties of these systems
strongly depend on the non-magnetic 3\textit{d} (X) and
\textit{sp} (Z) atoms. Early measurements by Webster et. al.,  on
several quaternary Heusler alloys as well as recent studies of
Walle \emph{et. al.}, on AuMnSn$_{1-x}$Sb$_{x}$ demonstrated the
importance of the \emph{sp} electrons in establishing the magnetic
properties.\cite{LB,AuMnSnSb} On the other hand, the importance of
the non-magnetic 3\textit{d} atoms for the magnetism of Heusler
alloys has been revealed recently by the experimental studies of
Duong \emph{et. al.}, and Ren \emph{et.
al.}\cite{CoCuMnSb,NiCuMnSb} The authors have shown the
possibility of tuning the Curie temperature of
Co$_{1-x}$Cu$_{x}$MnSb and Ni$_{1-x}$Cu$_{x}$MnSb alloys by the
substitution of Cu for Co and Ni, respectively. Furthermore, a
phase transition from a ferromagnetic to an antiferromagnetic
state is detected in both systems close to the stoichiometric
composition ($x\sim 1$). To reveal the nature of the magnetism in
the Mn-based Heusler alloys \c{S}a\c{s}{\i}o\u{g}lu \emph{et.
al.,} performed systematic first-principles calculations focusing
on the influence of the \emph{sp}-electrons on the magnetic
characteristics. The authors interpreted the obtained results
using the Anderson \emph{s-d} mixing model\cite{Anderson} and
showed that the complex magnetic behavior of the Mn-based Heusler
alloys can be described in terms of the competition of two
exchange mechanisms: the ferromagnetic RKKY-like exchange and the
antiferromagnetic superexchange.

Purpose of the given work is to investigate the influence of the
non-magnetic 3\emph{d} atoms on the magnetic properties of the
half-metallic semi-Heusler alloys: Co$_{1-x}$Cu$_{x}$MnSb and
Ni$_{1-x}$Cu$_{x}$MnSb. We determine the  magnetic phase diagram
of both systems at zero temperature. In agrement with the
experiments we obtain a phase transition from a ferromagnetic to
an antiferromagnetic state around $x\simeq 0.8$ and $x\simeq 0.6$
for the Co-based and Ni-based systems, respectively. The physical
mechanisms behind the magnetic phase transition is revealed. The
electronic structure calculations are performed using the
full-potential non-orthogonal local-orbital minimum-basis band
structure scheme (FPLO).\cite{fplo}

\begin{figure}[!t]
\begin{center}
\includegraphics[scale=0.52]{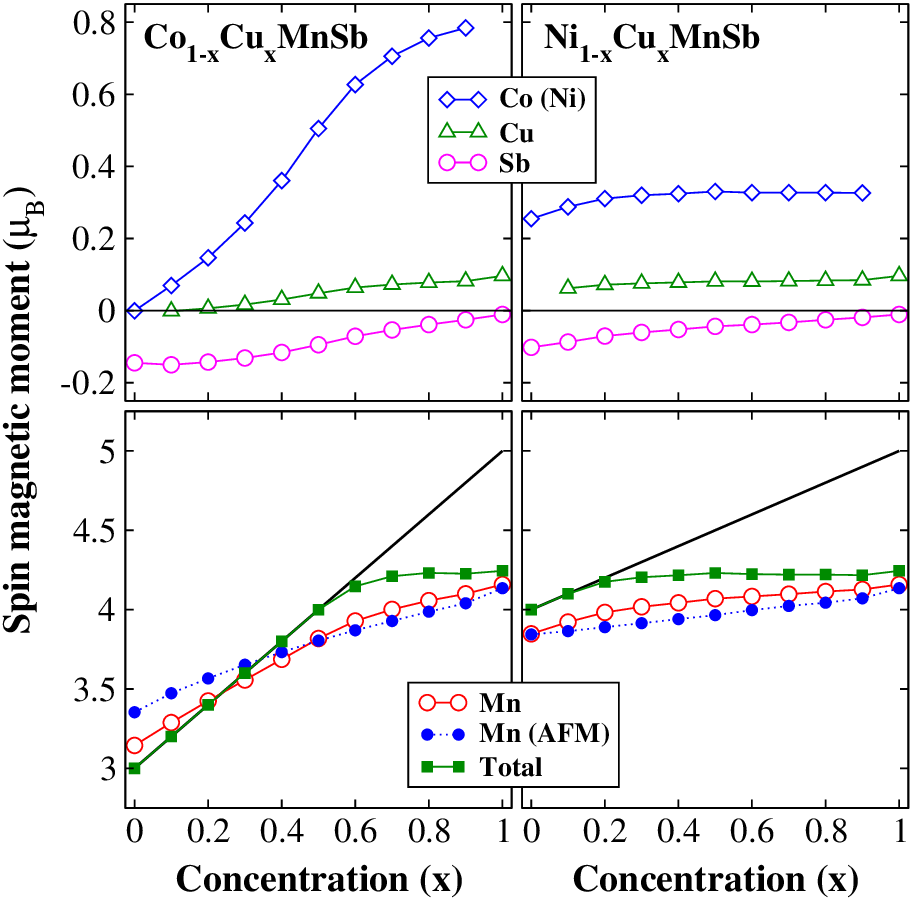}
\end{center}
\vspace*{-0.2cm} \caption{ (Color online) Calculated atom-resolved
and total spin moments (in $\mu_\mathrm{B}$) in
Co$_{1-x}$Cu$_{x}$MnSb and Ni$_{1-x}$Cu$_{x}$MnSb as a function of
the Cu concentration ($x$). The atom-resolved spin moments for
Co(Ni) and Cu have been scaled to one atom. Solid lines represent
the Slater-Pauling rule.\cite{GalanakisFull}} \label{fig1}
\end{figure}

We begin our discussion from the calculated magnetic moments.
Fig.~\ref{fig1} presents the  atom resolved and total magnetic
moments in Co$_{1-x}$Cu$_{x}$MnSb and Ni$_{1-x}$Cu$_{x}$MnSb  as a
function of the Cu content for the ferromagnetic state. For
comparison the Mn magnetic moment corresponding to the
antiferromagnetic state is given. As seen from Fig.~\ref{fig1},
for $x=0$ the corresponding parent compounds are half-metallic
with total integer magnetic moments of $3\mu_B$ and $4\mu_B$ for
CoMnSb and NiMnSb, respectively. As the Cu concentration increases
the total spin magnetic moment follows the Slater-Pauling  rule up
to $x\simeq 0.6$ ($x\simeq 0.2$) for the Co-based (Ni-based)
system and then it becomes almost constant. Thus the
half-metallicity is retained up to these particular values of the
Cu concentration. This can also be seen from the total density of
states (DOS) shown in Fig.~\ref{fig3} where the Fermi level cross
the spin minority states for the corresponding values of $x$.
Furthermore, the variation of the total magnetic moment is around
$1.25 \mu_B$ in the Co-based systems which mainly comes from the
Mn and Co atoms, whereas  this is rather small ($\simeq 0.25
\mu_B$) in Ni-based systems. The behavior of the induced moments
in Cu and Sb atoms only weakly depends on the $x$ concentration.
It should be noted that as seen from Fig.~\ref{fig1} the Mn moment
is insensitive to the magnetic order revealing the localized
nature of magnetism in HM semi-Heusler alloys and justifying the
use of Anderson \textit{s-d} model in the interpretation of the
results obtained from first-principles.

\begin{figure}[t]
\begin{center}
\includegraphics[scale=0.51]{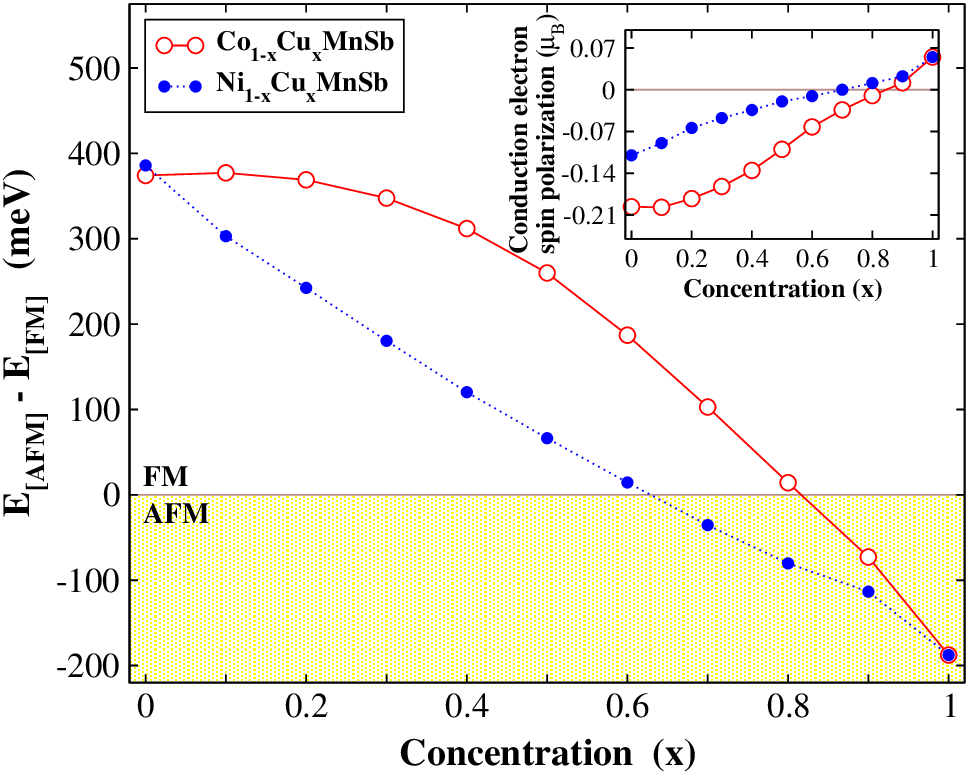}
\end{center}
\vspace*{-0.4cm} \caption{ (Color online) Ground state magnetic
phase diagram and total energy differences between AFM and FM
configurations of the Mn magnetic moments in
Co$_{1-x}$Cu$_{x}$MnSb  and Ni$_{1-x}$Cu$_{x}$MnSb as  a function
of the Cu concentration ($x$). In the inset we show the total spin
polarization of the  conduction electrons of X (Co, Ni, Cu) and Z
(Sb) atoms as a function of the Cu content.} \label{fig2}
\end{figure}

In order to confirm the  experimentally observed magnetic phase
transition we calculate the total energies corresponding to the FM
and the AFM  configurations of the  Mn magnetic moments. The zero
temperature  magnetic phase diagram is determined as the
difference of the corresponding total energies ($E_{AFM}-E_{FM}$)
and is presented in Fig.~\ref{fig2}. In agreement with the
experiments for both compounds we obtain a phase transition from a
ferromagnetic state to an antiferromagnetic one at a certain value
of the Cu concentration, $x$. As seen from Fig.~\ref{fig2} the
critical $x$ value for the Ni-based alloys ($x\simeq 0.6$) is
somehow smaller than the experimental value ($0.9 <x<1$). While in
the Co-based compounds the transition point ($x\simeq 0.8$) is
closer to the measured value ($0.9 <x<1$). Note that the HM
character, as discussed above, is lost before reaching the
transition point and the Fermi level crosses the minority-spin
conduction band but the ferromagnetism persists up to the
transition point.

As shown in Refs.~\onlinecite{Sasioglu2006} and
\onlinecite{Sasioglu2008}  the observed magnetic phase transition
in these systems can be qualitatively accounted for in terms of
the competition of the ferromagnetic RKKY-like exchange and
antiferromagnetic superexchange. Note that a detailed discussion
of the exchange mechanism in local moment systems  and
applications to different systems can be found  in Refs.
\onlinecite{Falicov} and \onlinecite{Levy}. Here we will give the
expressions for both exchange couplings in $\textbf{q} \rightarrow
0$ limit for the analysis of the obtained results. For
$\textbf{q}=0$ he RKKY-like coupling takes a simple form which is
$J_{\textrm{RKKY}}(0)=V^4D(\epsilon_{F})/E^{2}_h$, where $V$ is
the coupling between the Mn 3\textit{d} levels  and the conduction
electron states. The mixing interaction $V$ induces a spin
polarization in the conduction electron sea, and the propagation
of this polarization gives rise to an effective indirect exchange
coupling between distant magnetic moments. $D(\epsilon_{F})$ is
the density of states at the Fermi level and $E_h$ is the energy
required to promote an electron from the occupied 3\textit{d}
levels to the Fermi level. The value of the spin polarization of
the conduction electrons can be used to estimate the relative
contribution of this coupling. On the other hand, the
superexchange coupling does not posses a simple limit; for
$\textbf{q}=0$ it becomes
$J_{\textrm{S}}(0)=V^4\sum_{nk}[\epsilon_F-\epsilon_{nk}-E_h]^{-3}$,
where the sum is taken over the unoccupied states and the terms in
this sum drop off quickly as $\epsilon_{nk}$ increases. Thus, the
structure of the DOS above the Fermi level plays a key role in
determining the strength of this coupling.

\begin{figure}[t]
\begin{center}
\includegraphics[scale=0.52]{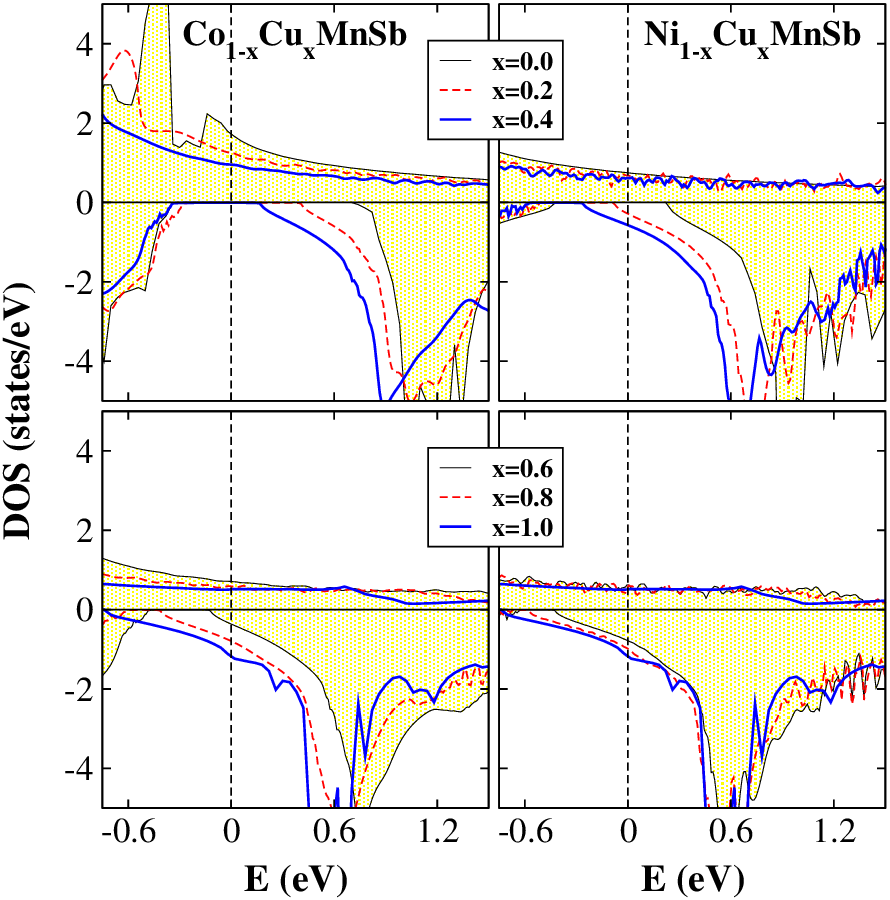}
\end{center}
\vspace*{-0.6cm} \caption{(Color online) Spin-resolved total
density of states (DOS) of Co$_{1-x}$Cu$_{x}$MnSb  and
Ni$_{1-x}$Cu$_{x}$MnSb around the Fermi level for selected values
of $x$. Vertical dotted lines denote the Fermi level. Positive
values of DOS correspond to the majority-spin electrons and
negative values to the minority-spin electrons.} \label{fig3}
\end{figure}

Now we return back to the discussion of the phase diagram in terms
of these two mechanisms. A qualitative information on the
variation of the RKKY-like  and superexchange contributions can be
obtained from the analysis of the conduction electron spin
polarization and from the structure of the DOS above the Fermi
level. As seen in Figs.~\ref{fig2} and \ref{fig3} when we
substitute Cu for Co(Ni), the spin polarization decreases and at
the same time the position of the Fermi level moves towards higher
energies, i.e., the number of the states just above the Fermi
level increases. This gives rise to an opposite behavior in the
relative contributions of the exchange mechanisms: a decrease for
the RKKY-like coupling and an increase in the superexchange
mechanism. In large part of the phase diagram the former coupling
is dominating. This is reflected as a correlation between the spin
polarization and the total energy differences given in
Fig.~\ref{fig2}. On the other hand, the superexchange coupling
becomes important for larger values of the Cu concentration, i.e.,
for $x>0.5$. As seen from Fig.~\ref{fig2} at the transition points
both mechanisms compensate each other giving rise to a spin
glass-like behavior.\cite{Turek} Further increase of $x$ leads to
an antiferromagnetic order in both compounds due to the dominating
character of the superexchange mechanism.

As discussed above the magnetic interactions in HM semi-Heusler
alloys are sensitive to the width of the gap and the position of
the Fermi level within the gap. Systems having large HM gaps and a
Fermi level far from the right edge of the gap are strongly
ferromagnetic and posses very high Curie
temperatures.\cite{Kubler,Sakuma,Sanyal,Rusz} This is due to the
fact that in this case the superexchange mechanism is less
efficient since the gap in the spin-down channel decreases the
number of available minority-spin states just above the Fermi
level. Thus, the position of the Fermi level within the gap is an
important parameter in determining the magnetic characteristics of
the HM ferromagnets. These findings suggest a way for tuning the
magnetic properties of the HM ferromagnets and allow the
fabrication of materials with predefined characteristics. It
should be noted that, as shown in Ref.~\onlinecite{Sasioglu2006},
the variation of the \textit{sp}-electrons (Z atom) concentration
is an alternative route for tuning the magnetic properties of the
Heusler alloys. However, in HM compounds both kind of atoms give
rise to similar effects as demonstrated by recent
experiments.\cite{AuMnSnSb,NiCuMnSb}

In conclusion, we study the effect of the non-magnetic 3\textit{d}
atoms on the magnetic properties of the half-metallic  Mn-based
semi Heusler alloys Co$_{1-x}$Cu$_{x}$MnSb and
Ni$_{1-x}$Cu$_{x}$MnSb ($0 \leq x \leq 1$) within the framework of
the parameter-free density functional theory. We show that the
magnetic interactions in these systems strongly depend on the
position of the Fermi level within the gap. We show that for Cu
concentrations preserving the half-metallic character the
ferromagnetic RKKY-like exchange mechanism is dominating, while
the antiferromagnetic superexchange coupling becomes important for
larger Cu concentrations and it is responsible for the observed
magnetic phase transition in both compounds. These findings can be
used as a practical tool to design materials with given physical
properties.


\begin{thebibliography}{99}

\bibitem{GalanakisFull}
I. Galanakis, P. H. Dederichs, and N. Papanikolaou, Phys. Rev. B
\textbf{66}, 174429 (2002).

\bibitem{LB}
P. J. Webster and K. R. A. Ziebeck, in {\em Alloys and Compounds
of d-Elements with Main Group Elements. Part 2.}, edited by H. R.
J. Wijn, Landolt-Bo\"ornstein, New Series, Group III, Vol. 19/c
(Springer-Verlag, Berlin 1988).


\bibitem{AuMnSnSb}
C. Walle, L. Offernes, and A. Kjekshus, J. Alloys and Comp.
\textbf{349}, 105 (2003).

\bibitem{CoCuMnSb}
N.~P. Duong, L.~T. Hung, T.~D. Hien, N.~P. Thuy, N.~T. Trung, and
E. Br\"{u}ck, J. Magn. Magn. Mater. \textbf{311}, 605 (2007).


\bibitem{NiCuMnSb}
S.~K. Ren, W.~Q. Zou, J. Gao, X.~L. Jiang, F.~M. Zhang, and Y.~W.
Du, J. Mag. Magn. Mater. \textbf{288}, 276 (2005).


\bibitem{Sasioglu2006}
E. \c{S}a\c{s}{\i}o\u{g}lu, L.~M. Sandratskii, and P. Bruno, Appl.
Phys. Lett. \textbf{89}, 222508 (2006).

\bibitem{Sasioglu2008}
E. \c{S}a\c{s}{\i}o\u{g}lu, L.~M. Sandratskii, and P. Bruno,
arXiv:0712.0158

\bibitem{Anderson}
P.~W. Anderson, Phys. Rev.  \textbf{124}, 41 (1961).


\bibitem{fplo}
K. Koepernik and H. Eschrig, Phys. Rev. B \textbf{59}, 3174
(1999); K. Koepernik, B. Velicky, R. Hayn, and H. Eschrig, Phys.
Rev. B \textbf{58}, 6944 (1998).


\bibitem{Falicov}
C.~E.~T. Gon\c{c}alves Da Silva and L.~M. Falicov, J. Phys. C:
Solid State Phys. \textbf{5}, 63 (1972).


\bibitem{Levy}
Z.~-P. Shi, P.~M. Levy, and J.~L. Fry,  Phys. Rev. B \textbf{49},
15159 (1994).


\bibitem{Turek}
P. Ferriani, I. Turek, S.  Heinze, G. Bihlmayer, and S. Bl\"{u}gel
Phys. Rev. Lett. \textbf{99}, 187203 (2007).


\bibitem{Kubler}
J. K\"{u}bler, Phys. Rev. B \textbf{67}, 220403 (2003).


\bibitem{Sakuma}
A. Sakuma, J. Phys. Soc. Jpn. \textbf{71}, 2534 (2002).

\bibitem{Sanyal}
B. Sanyal, L. Bergqvist, and O. Eriksson, Phys. Rev. B
\textbf{68}, 054417 (2003).


\bibitem{Rusz}
J. Rusz, L. Bergqvist, J. Kudrnovsk\'{y}, and I. Turek,  Phys.
Rev. B \textbf{73}, 214412 (2006).

\end{thebibliography}
\end{document}